\documentclass[pre,aps,floats,twocolumn,showpacs]{revtex4}
\usepackage[dvips]{epsfig}

\begin{document}

\title{Thermodynamics of two lattice ice models in three dimensions}

\author{Chizuru Muguruma$^{\rm\,a,b}$, Yuko Okamoto$^{\rm\,c}$, and
Bernd A. Berg$^{\rm \,d,e}$\footnote{Corresponding author.}}

\affiliation{ 
$^{\rm \,a)}$ Faculty of Liberal Arts, Chukyo University,
Toyoto, Aichi 470-0393, Japan\\
$^{\rm \,b)}$ School of Computational Science,
Florida State University, Tallahassee, FL 32306-4120, USA\\
$^{\rm \,c)}$ Department of Physics, Nagoya University,
Nagoya, Aichi 464-8602, Japan\\
$^{\rm \,d)}$ Department of Physics, Florida State
University, Tallahassee, FL 32306-4350, USA\\
$^{\rm \,e)}$ Institut f\"ur Theoretische Physik und 
Naturwissenschaflich-Theoretisches Zentrum (NTZ),
Universit\"at Leipzig, D-04009 Leipzig, Germany}

\date{\medskip \today }

\begin{abstract}
In a recent paper we introduced two Potts-like models in three
dimensions, which share the following properties: (A)~One of the ice 
rules is always fulfilled (in particular also at infinite temperature, 
$\beta =0$). (B)~Both ice rules hold for groundstate configurations.
This allowed for an efficient calculation of the residual entropy
of ice~I (ordinary ice) by means of multicanonical simulations. 
Here we present the thermodynamics of these models. Despite their 
similarities with Potts models, no sign of a disorder-order phase 
transition is found. 
\end{abstract}
\pacs{PACS: 05.50.+q, 11.15.Ha, 12.38.Gc, 25.75.-q, 25.75.Nq}
\maketitle

\section{Introduction}

By experimental discovery \cite{Gi33} it was found that ice~I (ordinary 
ice) has in the zero temperature limit a residual entropy $S/N = k\,
\ln(W_1)>0$ where $N$ is the number of molecules and $W_1$ the number 
of configurations per molecule. Subsequently Linus Pauling \cite{Pa35} 
based the estimate $W_1^{\rm Pauling}=3/2$ on the ice rules:
\begin{enumerate}
\item There is one hydrogen atom on each bond (then called hydrogen 
      bond). 
\item There are two hydrogen atoms near each oxygen atom (these three 
      atoms constitute a water molecule). 
\end{enumerate}
Pauling's combinatorial estimate turned out to be in excellent agreement 
with subsequent refined experimental measurements~\cite{Gi36}. This may
be a reason why it took 25 years until Onsager and Dupuis \cite{OnDu60} 
pointed out that $W_1=1.5$ is only a lower bound. Subsequently Nagle 
\cite{Na65} used a series expansion method to derive the estimate 
$W_1^{\rm Nagle}=1.50685\,(15)$. 

\begin{figure}[-t] \begin{center}
\epsfig{figure=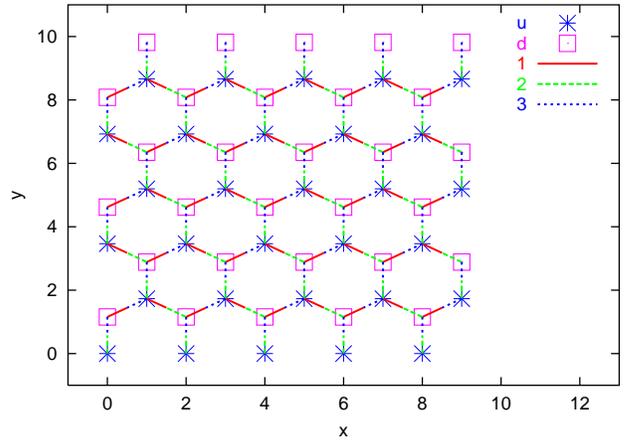,width=\columnwidth} 
\caption{Lattice structure of one layer of ice~I (reproduced from
Ref.~[7]). The up (u) sites are at $z=1/\sqrt{24}$ and the down 
(d) sites at $z=-1/\sqrt{24}$.  Three of its four pointers to nearest 
neighbor sites are shown.} \label{fig_icepnt}
\end{center} \end{figure} 

In \cite{Be07} we introduced two simple models with nearest neighbor 
interactions on 3D hexagonal lattices, which allow one to calculate 
the residual entropy of ice~I by means of multicanonical (MUCA) 
\cite{Be92a,Be92b,BBook} simulations. The hexagonal lattice structure
is depicted in Fig.~\ref{fig_icepnt}~\cite{fig01}. 
In the first model, called 6-state (6-s) H$_2$O molecule model, ice 
rule~(2) is always enforced and we allow for six distinct orientations 
of each H$_2$O molecule. Its energy is defined by
\begin{equation} \label{E1}
  E = - \sum_b h(b,s^1_b,s^2_b)\ .
\end{equation}
Here, the sum is over all bonds $b$ of the lattice ($s^1_b$ and 
$s^2_b$ indicate the dependence on the states of the two H$_2$O 
molecules, which are connected by the bond) and
\begin{equation} \label{hs}
  h(b,s^1_b,s^2_b) = 
  \cases{ 1\ {\rm for\ a\ hydrogen\ bond}\,, \cr
               0\ {\rm otherwise}\,.  }
\end{equation}
In the second model, called 2-state (2-s) H-bond model, ice rule~(1) 
is always enforced and we allow for two positions of each hydrogen 
nucleus on a bond. The energy is defined by
\begin{equation} \label{E2}
  E = - \sum_s f(s,b^1_s,b^2_s,b^3_s,b^4_s)\,,
\end{equation}
where the sum is over all sites (oxygen atoms) of the lattice and
$f$ is given by
\begin{eqnarray} \label{fs}
 & f(s,b^1_s,b^2_s,b^3_s,b^4_s)\ = \qquad & \\ \nonumber 
  & \cases{ 
  2\ {\rm for\ two\ hydrogen\ nuclei\ close\ to}\ s\,, \cr 
  1\ {\rm for\ one\ or\ three\ hydrogen\ nuclei\ close\ to}\ s\,,\cr 
  0\ {\rm for\ zero\ or\ four\ hydrogen\ nuclei\ close\ to}\ s\,. } &
\end{eqnarray}
The groundstates of either model fulfill both ice rules. 

In this paper we use units with $k=1$ for the Boltzmann constant, i.e., 
$\beta=1/T$. At $\beta=0$ the number of configurations is $6^N$ for the 
6-s model and $2^{2N}$ for the 2-s model. This sets the normalization, 
which can then be connected by a MUCA simulation of the type 
\cite{Be92b} to $\beta$ values large enough so that groundstates 
get sampled. In reasonably good agreement with Nagle the estimate 
$W_1^{\rm MUCA} = 1.50738\,(16)$ was obtained in \cite{Be07}. In 
Ref.~\cite{BW07} these calculation were extended to partially ordered 
ice for which corrections to groundstate entropy estimates by Pauling's 
method were previously not available.

A considerable literature 
\cite{YaNa79,NaKr91,KrNa92,Na93,BaNe98,Wa99,Li05,Gi06} exists on 
lattice ice models. Most of these papers deal with 2D square ice. 
An extension to 3D is considered in \cite{NaKr91,KrNa92,Gi06}. 
All these models have in common that they enforce both ice rules 
generically and not just for the groundstates. So, they are 
non-trivial at all coupling constant values, while it is precisely 
the triviality of our ice models at $\beta =0$, which allows one to 
set the normalization for the entropy and free energy, and they share 
with certain spin models \cite{ChWu87} that the residual entropy of 
their groundstates violates the third law of thermodynamics.

Superficially our models are similar to $q$-state Potts models 
\cite{Wu82} with $q=6$ and the Ising case $q=2$, which have first
($q=6$) and second ($q=2$) order phase transitions in 2D as well as 
in 3D. In contrast to that, we provide numerical evidence in this 
paper that our ice models do not undergo a disorder-order phase 
transition at any finite value of $\beta$. Our results are presented 
in section~\ref{sec_sim}. Summary and conclusions follow in 
section~\ref{sec_sum}.

\section{Simulation results\label{sec_sim}}

\begin{table}[ht] 
\caption{Overview of our multicanonical simulations.
Here, $n_x$, $n_y$, $n_z$ are the number of lattice sites
along the $x$, $y$, $z$ axes, and $N=n_x\,n_y\,n_z$. } 
\label{tab_stat}
\medskip \centering
\begin{tabular}{|c|c|c|c|c|c|}   \hline
$n_x$&$n_y$&$n_z$& $N$ & cycles (6-s) & cycles (2-s) \\ \hline
   4 &   8 &  4  & 128 & 37828 & 141825\\ \hline
   4 &  12 &  6  & 288 &  9455 &  33205\\ \hline
   5 &  12 &  6  & 360 &  4891 &  21621\\ \hline
   6 &  12 &  8  & 576 &   653 &  11479\\ \hline
   7 &  16 &  8  & 896 &   412 &   6452\\ \hline
   8 &  20 & 10  &1600 &   215 &   1587\\ \hline
  10 &  24 & 12  &2880 &  1133 &    506\\ \hline
\end{tabular} \end{table}


Using periodic boundary conditions (BCs), our simulations are based 
on a lattice construction \cite{Be05} similar to that set up for Potts 
models in \cite{BBook}. The lattice sizes used are compiled in 
table~\ref{tab_stat}.  The lattice contains then $N=n_x\,n_y\,n_z$ 
sites, where $n_x$, $n_y$, and $n_z$ are the number of sites along 
the $x$, $y$, and $z$ axes, respectively. Periodic BCs restrict the 
allowed values of $n_x$, $n_y$, and $n_z$ to $n_x = 1,\,2,\,3,\,\dots$, 
$n_y = 4,\,8, \,12,\,\dots$, and $n_z = 2,\,4,\,6, \,\dots$~. Otherwise 
the geometry does not close properly. 

As proposed in \cite{Be03} we used a Wang-Landau 
\cite{WL01} recursion for determining the  MUCA weights and performed
subsequent MUCA data production with fixed weights. With one exception 
we used $32\times (20\times 10^6)$ sweeps per lattice for data 
production. For the largest lattice of the 6-s model we produced a 
ten times larger statistics. Table~\ref{tab_stat} listed for each 
lattice size and model the number of cycling events from the average 
disordered energy $E_0$ at $\beta=0$ to the groundstate energy $E_g$ 
and back,
\begin{equation} 
  E_0~~\leftrightarrow~~E_g\,,
\end{equation} 
as recorded during the production part of the run. From the energy 
functions (\ref{E1}) and (\ref{E2}) one finds $E_0=-N$ for the 6-s 
model (there are two hydrogen atoms per oxygen and the probability
to form a hydrogen bond is 1/2), $E_0=-1.25\,N$ for the 2-s model (at 
one site there are 16 arrangements of hydrogen atoms with average 
energy contribution $-[2\times 0 +8\times 1 +6\times 2]/16=-1.25$), 
and $E_g= -2N$ for both models. In the following we restrict the
$\beta$ range of our figures to $0\le \beta\le 5$, which is large 
enough to sample groundstates in sufficient numbers so that 
extrapolations down to temperature $T=0$ become controlled.

\begin{figure}[-t] \begin{center}
\epsfig{figure=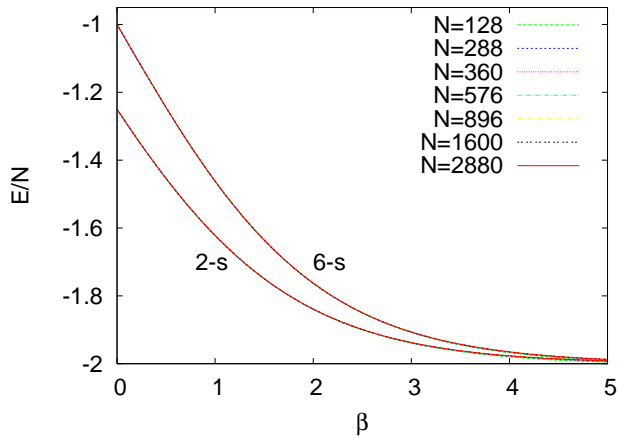,width=\columnwidth} 
\caption{Energy per site for the 6-s and 2-s models.} \label{fig_e}
\end{center} \end{figure} 

\begin{figure}[-t] \begin{center}
\epsfig{figure=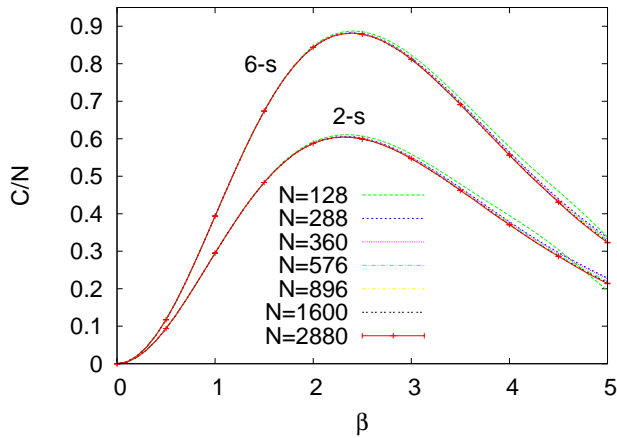,width=\columnwidth} 
\caption{Specific heat per site for the 6-s and 2-s models.} 
\label{fig_c}
\end{center} \end{figure} 

In Fig.~\ref{fig_e} we show the average energy per site, $E/N$, from 
the MUCA simulations of our two models as obtained by the reweighting 
procedure~\cite{BBook} (note that we use $E$ for the energy of 
configuration as well as for average values over configuration 
energies and assume the reader knows to distinguish them). Obviously 
there are almost no finite size effects, because the curves from all 
lattice sizes fall within small statistical errors, which are not visible
on the scale of this figure, on top of one another. 

\begin{table}[ht]
\caption{Some specific heat data $C/N$ with error bars 
(in parenthesis) for the $N=2880$ lattice.} \label{tab_c}
\medskip \centering
\begin{tabular}{|c|c|c|}   \hline
$\beta$ & 6-s model      & 2-s model     \\ \hline
  0.5   & 0.1175119 (77) & 0.093780 (17) \\ \hline
  1.5   & 0.673681 (71)  & 0.48331 (11)  \\ \hline
  2.5   & 0.87873 (19)   & 0.59913 (22)  \\ \hline
  3.5   & 0.69110 (35)   & 0.46235 (42)  \\ \hline
  4.5   & 0.43066 (45)   & 0.28637 (61)  \\ \hline
\end{tabular} \end{table}

The specific heat per site, $C/N$, is calculated via the 
fluctuation-dissipation theorem,
\begin{equation} 
  C = \frac{dE}{dT} = - \beta^2\,\frac{dE}{d\beta} = \beta^2\,
      \left( \langle E^2\rangle - \langle E \rangle^2 \right)\,,
\end{equation} 
and plotted in Fig.~\ref{fig_c}. Finite size corrections are now
visible for the smallest, $N=128$, lattice. For the other lattices
the curves fall within error bars on top of one another. Error bars 
were calculated with respect to 32 jackknife bins and are at some
$\beta$ values included for our largest, $N=2880$, lattice. Some data 
for these points are given in table~\ref{tab_c}. Note that the $N=2880$ 
data for the 6-s model rely on a ten times large statistics than those 
for the 2-s model, while the error bars are only slightly smaller. As 
noticed before \cite{Be07}, the simulations of the 2-s model are more
efficient for determining the groundstate entropy than simulations of
the 6-s model. Fluctuations increase with lattice size, so that it
is more difficult to obtain accurate results on large than on small
lattices.

We want to contrast Fig.~\ref{fig_c} with specific heat results for the 
6-state and 2-state Potts models on $L^D$ lattices. Immediately, one 
notices that it is not entirely clear whether this comparison should be 
done in 2D or 3D. While the space dimension in which our ice models are 
embedded is clearly 3D, each site is connected through links with four 
neighboring sites, which is the Potts model situation in 2D. The 2D and 
3D Ising models are well known for their second order phase transitions. 
The specific heat is logarithmically divergent in 2D \cite{On44} and 
has a critical exponent $\alpha\approx 0.1$ in 3D (see \cite{PV02} 
for a review). The 2D and 3D 6-state Potts models have first order 
transitions with a larger latent heat per spin in 3D than in 2D (in 
the normalization of \cite{BBook} $\triangle E/N = 0.40292828$ in 2D 
\cite{Ba73} and $\triangle E/N = 2.36442\pm 0.00017$ in 3D \cite{BBD08}).

For second order transitions the maximum of the specific heat diverges 
$\sim \ln(L)$ for a logarithmic divergence ($\alpha=0$) and $\sim 
L^{\alpha/\nu}$ for $\alpha>0$, where $\nu$ is the critical exponent 
of the correlation length. In case of first order phase transitions 
the peak in the specific heat diverges $\sim L^D$, where the 
proportionality factor is \cite{CLB86} $(\beta_t)^2(\triangle E/N)^2$ 
with $\beta_t$ the inverse transition temperature and $\triangle E/N$
the latent heat per spin. 

\begin{figure}[-t] \begin{center}
\epsfig{figure=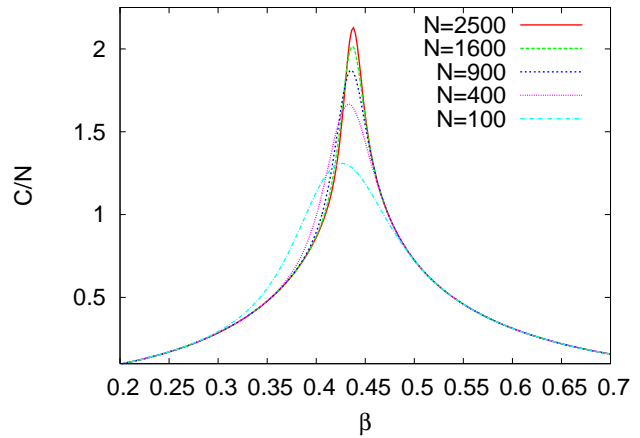,width=\columnwidth} 
\caption{Specific heat per site for the 2D Ising model on $N=L^2$
lattices.} \label{fig_2DIC}
\end{center} \end{figure} 

\begin{figure}[-t] \begin{center}
\epsfig{figure=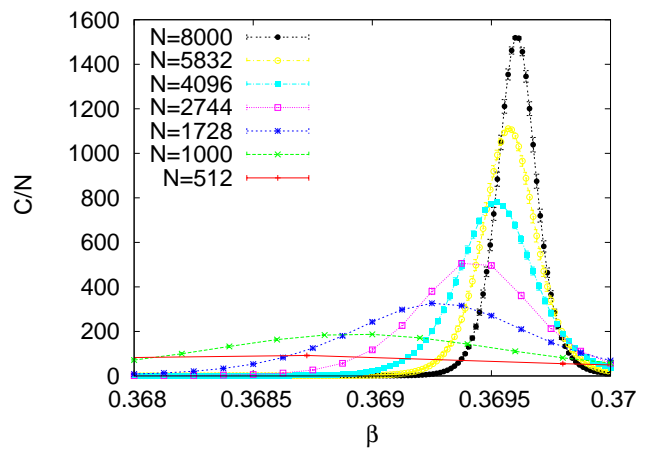,width=\columnwidth} 
\caption{Specific heat per site for the 3D 6-state Potts model 
on $N=L^3$ lattices.} \label{fig_3D6qC}
\end{center} \end{figure} 

In Figs.~\ref{fig_2DIC} and~\ref{fig_3D6qC} we plot the specific 
heat on various lattices for the two extremes, the weak logarithmic 
divergence for the 2D Ising model and the strong divergence for
the 3D 6-state Potts model. For the 2D Ising model the analytical
solutions of Ferdinand and Fisher~\cite{FF69} are plotted, while
the plots for the 3D 6-state Potts model rely on recent numerical 
results \cite{BBD08}. It is clear that even the case of a weak 
logarithmic divergence is markedly distinct from the behaviors in 
Fig.~\ref{fig_c}, where no finite size effects are observed within 
the rather accurate statistical errors. This distinction becomes
all too obvious when the comparison is made with the strong
first order phase transition of the 3D 6-state Potts model.

\begin{figure}[-t] \begin{center}
\epsfig{figure=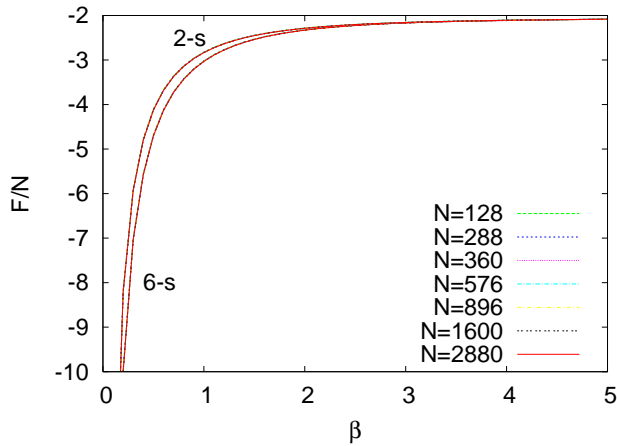,width=\columnwidth} 
\caption{Free energy . } \label{fig_fe}
\end{center} \end{figure} 

\begin{figure}[-t] \begin{center}
\epsfig{figure=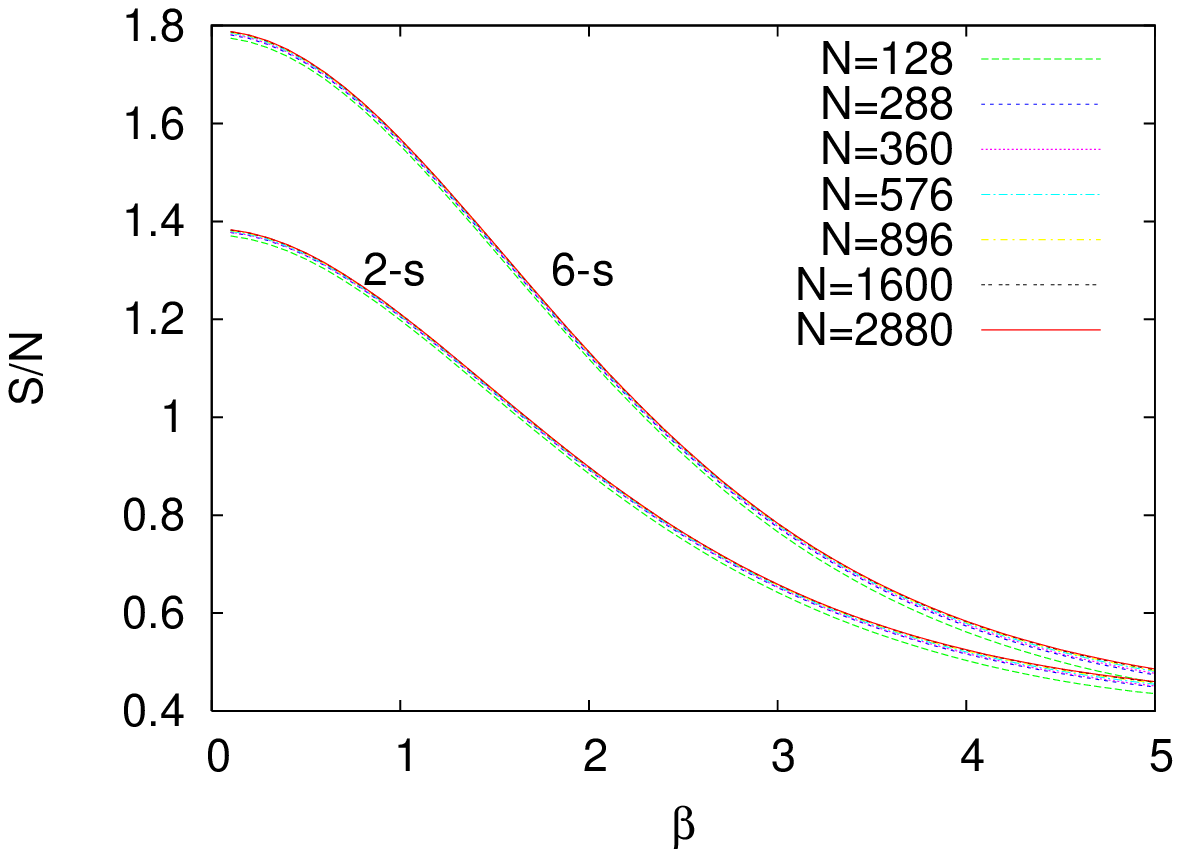,width=\columnwidth} 
\caption{Entropy . } \label{fig_s}
\end{center} \end{figure} 

To complete the picture of our two ice models we plot in 
Figs.~\ref{fig_fe} and~\ref{fig_s} their free energy and
entropy densities as obtained from our simulations, using as 
input the known normalizations at $\beta=0$. In
the cases at hand these are $S_0/N=\ln(6)$ for the 6-s and 
$S_0/N=\ln(4)$ for the 2-s model. Relative statistical errors
are smaller than those in Fig.~\ref{fig_c} for the specific 
heat. In the $\beta\to\infty$ limit our data improve slightly
on the results reported in Ref.~\cite{Be07}, because we have
with $N=2880$ one larger lattice added. Consistent fits to
the previously discussed form $W_1(x)=W_1^{\rm MUCA}+a_1x^{\theta}$,
$x=1/N$ combine to
\begin{equation} 
  W_1^{\rm MUCA}=1.50721\,(13)~~{\rm and}~~\theta=0.901\,(16)\ .
\end{equation} 
The error bars in parenthesis are purely statistical and do not
reflect eventual, additional systematic errors due to higher 
order finite size corrections.
\bigskip

\section{Summary and Conclusions \label{sec_sum}}

The unusual properties of water and ice owe their existence to a 
combination of strong directional polar interactions and a network 
of specifically arranged hydrogen bonds \cite{BeFo33,EiKa69,PeWh99}. 
The groundstate structure of such a network can be described by simple 
lattice models, which are defined by the energy functions (\ref{E1}) 
and (\ref{E2}).

In the present paper we have presented finite size scaling evidence 
that there is no phase transition between $\beta =0$ and the 
groundstate region of these models. This lack of a transition
makes reliable estimates of the combinatorial groundstate entropy 
of ice~I particularly easy.

Tentatively, we like to see a reason for the marked difference to $q=6$ 
and $q=2$ Potts models in the large groundstate entropy, $S/N=\ln(W_1)$, 
of our ice models, which violates the third law of thermodynamics, 
while the groundstate entropy of Potts models, $S/N = \ln(q)/N$, 
approaches zero in the $N\to\infty$ limit. This is not an entirely 
convincing argument as the effective number of states $W_1$ per spin 
is still about 2 (i.e., larger than 1.5) for the 3D 6-state Potts model 
at the ordered endpoint of the transition~\cite{BBD08}.

Note that we did not investigate bond statistics in the groundstate 
ensemble, which one may expect to exhibit critical correlations.

\acknowledgments We thank Santosh Dubey for providing the plot of 
Fig.~\ref{fig_3D6qC}. Partial support for this work was received
for BB from the JSPS and the Humboldt Foundation, for CM by the 
Chukyo University Research Fund, and for YO by the Ministry of 
Education, Culture, Sports, Science and Technology of Japan, 
Grants-in-Aid for the Next Generation Super Computing Project,
Nanoscience Program and for Scientific Research in Priority Areas, 
Water and Biomolecules. BB acknowledges useful discussions with
Wolfhard Janke.


\begin{thebibliography}{19}
\bibitem{fig01} The physical distances in the figure do not matter 
                in the present context. The interested reader can 
                find them explained in~\cite{Be07}).
\end{thebibliography}
\end{document}